\documentclass{elsart}

\usepackage{epsfig}

\begin{document}

\begin{frontmatter}

\title{A Study of the Personal Income Distribution in Australia}

\author[UMD]{Anand Banerjee},
\author[UMD]{Victor M. Yakovenko},
\author[ANU]{T. Di Matteo}

\address[UMD]{Department of Physics, University of Maryland, College
  Park, Maryland 20742-4111, USA}

\address[ANU]{Department of Applied Mathematics, The Australian
  National University, Canberra, ACT 0200, Australia}


\begin{abstract}
\small

We analyze the data on personal income distribution from the
Australian Bureau of Statistics.  We compare fits of the data to the
exponential, log-normal, and gamma distributions.  The exponential
function gives a good (albeit not perfect) description of 98\% of the
population in the lower part of the distribution.  The log-normal and
gamma functions do not improve the fit significantly, despite having
more parameters, and mimic the exponential function.  We find that the
probability density at zero income is not zero, which contradicts the
log-normal and gamma distributions, but is consistent with the
exponential one.  The high-resolution histogram of the probability
density shows a very sharp and narrow peak at low incomes, which we
interpret as the result of a government policy on income
redistribution.

\end{abstract}


\end{frontmatter}

\section{Introduction}
\label{}

The study of income distribution has a long history.  More than a
century ago, Pareto \cite{Pareto} proposed that income distribution
obeys a universal power law, valid for all time and countries.
Subsequent studies found that this conjecture applies only to the top
1$\div$3\% of the population.  The question of what is the
distribution for the majority (97$\div$99\%) of population with lower
incomes remains open.  Gibrat \cite{Gibrat} proposed that income
distribution is governed by a multiplicative random process resulting
in the log-normal distribution.  However, Kalecki \cite{Kalecki}
pointed out that such a log-normal distribution is not stationary,
because its width keeps increasing with time.  Nevertheless, the
log-normal function is widely used in literature to fit the lower part
of income distribution \cite{DiMatteo,Souma,Gallegati}.  Yakovenko and
Dr\u{a}gulescu \cite{DY-money} proposed that the distribution of
individual income should follow the exponential law analogous to the
Boltzmann-Gibbs distribution of energy in statistical physics.  They
found substantial evidence for this in the statistical data for USA
\cite{DY-income,DY-wealth,DY-survey,SY-evolution}.  Also widely used
is the gamma distribution, which differs from the exponential one by a
power-law prefactor \cite{Mimkes,West,Ferrero}.  For a recent
collection of papers discussing these distributions, see the book
\cite{Kolkata}.

Distribution of income $x$ is characterized by the probability density
function (PDF) $P(x)$, defined so that the probability to find income
in the interval from $x$ to $x+dx$ is equal to $P(x)\,dx$. The PDFs
for the distributions discussed above have the following functional
forms:
\begin{eqnarray}
  P(x) &=& \left\{
  \begin{array}{lll}
  {1\over T}\exp(-x/T)  &{}& \mbox{exponential,} \\
  {1\over xs\sqrt{2\pi}}\exp\left[{-\log^2(x/m) \over 2s^{2}}\right]
  &{}& \mbox{log-normal,}\\
  \frac{(\beta)^{-(1+\alpha)}}{\Gamma(1+\alpha,0)}x^\alpha
  \exp(-x/\beta) &{}& \mbox{gamma.}
  \end{array}\right.
\label{pdf}
\end{eqnarray}
The exponential distribution has one parameter $T$, and its $P(x)$ is
maximal at $x=0$.  The log-normal and gamma distributions have two
parameters each: $(m,s)$ and $(\beta,\alpha)$.  They have maxima
(called modes in mathematical statistics) at $x=m\e^{-s^{2}}$ and
$x=\alpha\beta$, and their $P(x)$ vanish at $x=0$.  Many researchers
impose the condition $P(0)=0$ \emph{a priori}, ``because people cannot
live on zero income''.  However, this assumption must be checked
against the real data.

In this paper, we analyze statistical data on personal income
distribution in Australia for 1989--2000 and compare them with the
three functions in Eq.\ (\ref{pdf}).  The data were collected by the
Australian Bureau of Statistics (ABS) using surveys of population. The
anonymous data sets give annual incomes of about 14,000 representative
individuals, and each individual is assigned a weight. The weights add
up to $1.3\div1.5\times10^7$ in the considered period, which is
comparable to the current population of Australia of about 20 million
people.  In the data analysis, we exclude individuals with negative
and zero income, whose total weight is about 7\%. These ABS data were
studied in the previous paper \cite{DiMatteo}, but without weights and
with the emphasis on the Pareto tail at high income.  Here we
re-analyze the data in the middle and low income range covering about
99\% of the population, but excluding the Pareto tail.  The number of
data points in the Pareto tail is relatively small in surveys of
population, which complicates accurate analysis of the tail.

\section{Cumulative Distribution Function}
\label{sec:CDF}

In this Section, we study the cumulative distribution function (CDF)
$C(x)=\int_x^\infty P(x')\,dx'$.  The advantage of CDF is that it can
be directly constructed from a data set without making subjective
choices. We sort incomes $x_n$ of $N$ individuals in decreasing order,
so that $n=1$ corresponds to the highest income, $n=2$ to the second
highest, etc. When the individuals are assigned the weights $w_n$, the
cumulative probability for a given $x_n$ is $C=\sum_{k=1}^n
w_k/\sum_{k=1}^N w_k$, i.e.\ $C(x)$ is equal to the normalized sum of
the weights of the individuals with incomes above $x$.  We fit the
empirically constructed $C(x)$ to the theoretical CDFs corresponding
to Eq.\ (\ref{pdf})
\begin{eqnarray}
  C(x) &=& \left\{
  \begin{array}{lll}
  \exp(-x/T)  &{}& \mbox{exponential}, \\
  \frac{1}{2}
  \left[1-{\rm Erf}\left({\log(x/m)\over s\sqrt{2}}\right)\right]
  &{}& \mbox{log-normal}, \\
  \Gamma(1+\alpha,x/\beta)/\Gamma(1+\alpha,0) &{}& \mbox{gamma},
  \end{array} \right.
\label{cdf}
\end{eqnarray}
where ${\rm Erf}(x)=\frac{2}{\sqrt{\pi}}\int_0^x e^{-z^{2}}\,dz$ is
the error function, and $\Gamma(\alpha,x)=\int_x^\infty
z^{\alpha-1}e^{-z}\,dz$.

\begin{figure}[b]
\includegraphics[width=0.52\linewidth]{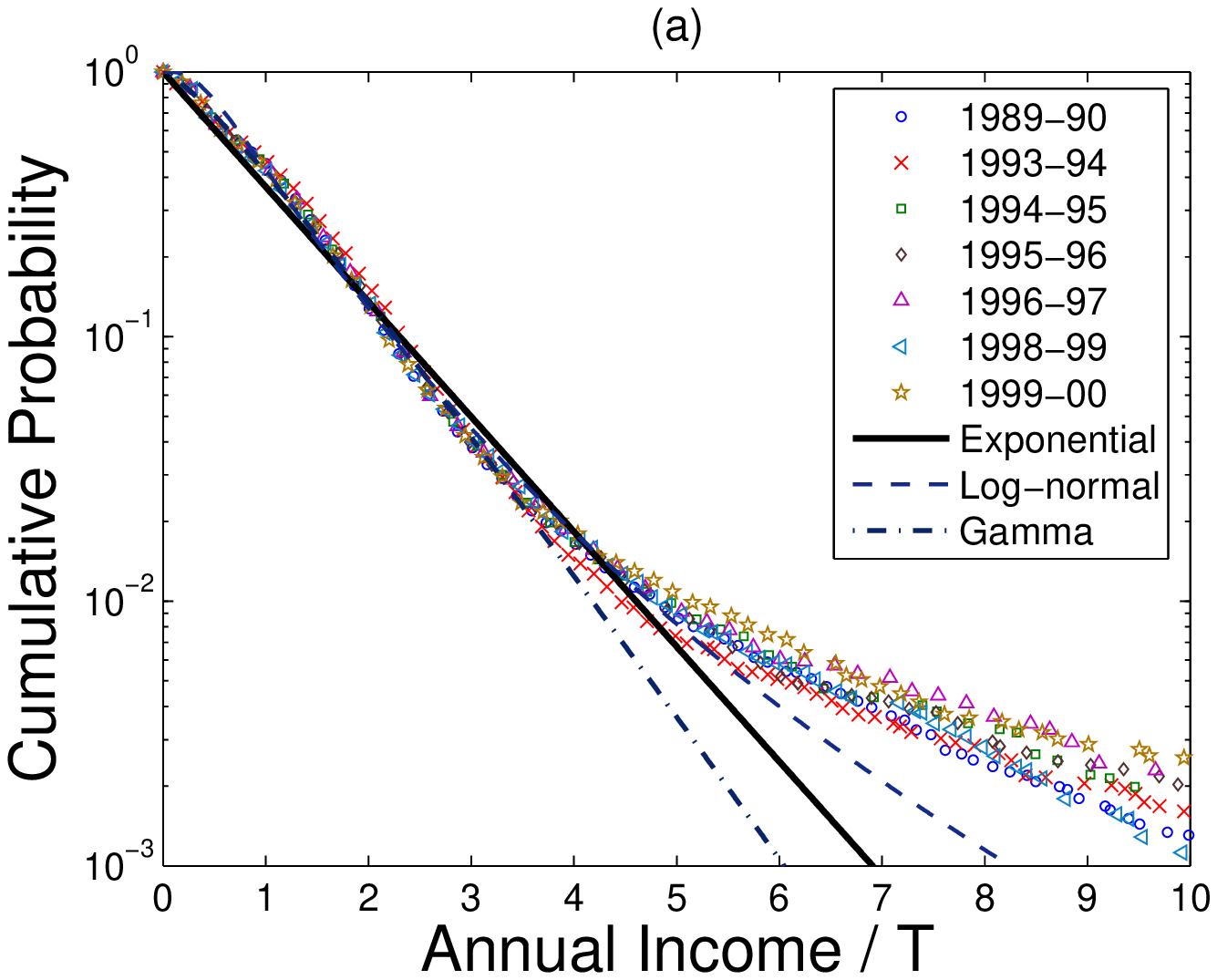}
\hfill
\includegraphics[width=0.52\linewidth]{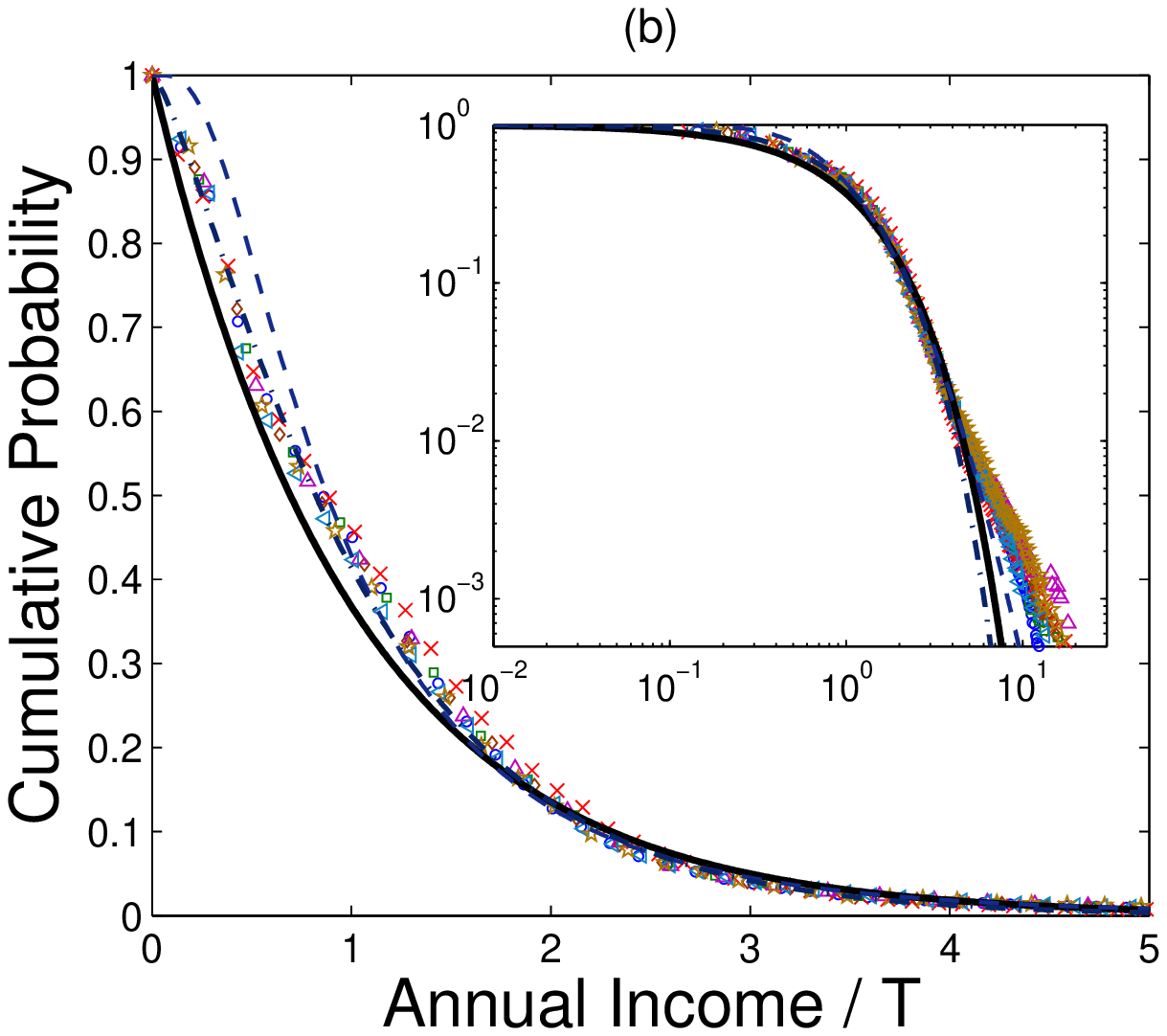}
\caption{The cumulative distribution function (CDF) of income,
  shown in the log-linear (a), linear-linear (b), and log-log (inset)
  scales.  The income values for different years are normalized to the
  parameter $T$ of the exponential distribution, given in Table
  \ref{table}.  The lines show fits to different theoretical
  distributions in Eq.\ (\ref{cdf}).}
\label{fig:CDF}
\end{figure}

To visualize $C(x)$, different scales can be used.  Fig.\ 
\ref{fig:CDF}(a) uses the log-linear scale, i.e. shows the plot of
$\ln C$ vs.\ $x$.  The main panel in Fig.\ \ref{fig:CDF}(b) uses the
linear-linear scale, and the inset the log-log scale, i.e.\ $\ln C$
vs.\ $\ln x$.  We observe that the log-linear scale is the most
informative, because the data points approximately fall on a straight
line for two orders of magnitudes, which suggests the exponential
distribution.  To obtain the best fit in the log-linear scale, we
minimize the relative mean square deviation $\sigma^2 =
\frac1M\sum_{i=1}^M\left(\frac{C_e(x_i)-C_t(x_i)}{C_e(x_i)}\right)^2
\approx\frac1M\sum_{i=1}^M\{\ln[C_e(x_i)]-\ln[C_t(x_i)]\}^2$ between
the empirical $C_e(x)$ and theoretical $C_t(x)$ CDFs.  For this sum,
we select $M=200$ income values $x_i$ uniformly spaced between $x=0$
and the income at which CDF is equal to 1\%, i.e. we fit the
distribution for 99\% of the population.  The minimization procedure
was implemented numerically in Matlab using the standard routines.

For the exponential distribution, the fitting parameter $T$ determines
the slope of $\ln C$ vs.\ $x$ and has the dimensionality of Australian
dollars per year, denoted as AUD or simply \$ (notice that $1\,{\rm
  k}\$=10^3\,\$ $).  $T$ is also equal to the average income $\langle
x\rangle$ for the exponential distribution.  The parameters $m$ and
$\beta$ for the log-normal and gamma distributions also have the
dimensionality of AUD, and the average incomes $\langle x\rangle$ for
these two distributions are $me^{s^2/2}$ and
$\beta\Gamma(\alpha+2,0)/\Gamma(\alpha+1,0)$.  The parameters $s$ and
$\alpha$ are dimensionless and characterize the shape of the
distributions. The values of these parameters, obtained by fits for
each year, are given in Table \ref{table}. Using the values of $T$, we
plot $C$ vs.\ $x/T$ in Fig.\ \ref{fig:CDF}.  In these coordinates, the
CDFs for different years (shown by different symbols) collapse on a
single curve for the lower 98\% of the population.  The collapse
implies that the shape of income distribution is very stable in time,
and only the scale parameter $T$ changes in nominal dollars.  The
three lines in Fig.\ \ref{fig:CDF} show the plots of the theoretical
CDFs given by Eq.\ (\ref{cdf}).  In these coordinates, the exponential
CDF is simply a straight line with the slope $-1$.  For the plots of
the log-normal and gamma CDFs, we used the parameters $\overline
s=0.72$, $\overline{m/T}=0.88$, $\overline\alpha=0.38$, and
$\overline{\beta/T}=0.77$ obtained by averaging of the parameters in
Table \ref{table} over the years.  We observe that all three
theoretical functions give reasonably good, albeit not perfect, fits
of the data with about the same quality, as confirmed by the values of
$\sigma$ in Table \ref{table}.  Although the log-normal and gamma
distributions have the extra parameters $s$ and $\alpha$, the fitting
procedure selects their values in such a way that these distributions
mimic the exponential shape.  Actually, we constructed the gamma fit
only for 98\% of the population, because the fit for 99\% gives
$\alpha=0$, i.e.\ the exponential.  We conclude that the exponential
distribution gives a reasonable fit of the empirical CDFs with only
one fitting parameter, whereas the log-normal and gamma distributions
with more fitting parameters do not improve the fit significantly and
simply mimic the exponential shape.

\begin{table}[b]
\caption{Parameters of the distributions (\ref{pdf}) and (\ref{cdf})
  obtained by minimization of the relative mean square deviation $\sigma^2$ 
  between the empirical and theoretical CDFs.  The last column gives 
  position of the sharp peak in Fig.\ \ref{fig:PDF}(b).}
\centering
\begin{tabular}{|c|c|c|c|c|c|c|c|c|c|}
\hline Year & T & m & s &  $\beta$ & $\alpha$ &
\multicolumn{3}{|c|}{$\sigma$ } & Peak \\
\cline{7-9} & k\$ & k\$ & & k\$ & & Exp & L-N & Gamma & \$ \\ \hline
1989-90 & $ 17.8 $ & $ 15.1 $ & 0.74 & $ 13.4 $ & 0.39 & $ 13\% $ & $ 11\%  $ &  $ 6.8\% $ & 6196 \\
1993-94 & $ 18.5 $ & $ 18.8 $ & 0.63 & $ 13.1 $ & 0.59 & $ 18\% $ & $ 9.6\% $ &  $ 5.7\% $ & 7020 \\
1994-95 & $ 19.6 $ & $ 17.7 $ & 0.71 & $ 14.9 $ & 0.40 & $ 15\% $ & $ 9.4\% $ &  $ 5.5\% $ & 7280 \\
1995-96 & $ 20.5 $ & $ 18.2 $ & 0.72 & $ 15.7 $ & 0.39 & $ 14\% $ & $ 8.6\% $ &  $ 6.5\% $ & 7280 \\
1996-97 & $ 21.2 $ & $ 18.9 $ & 0.72 & $ 16.5 $ & 0.37 & $ 14\% $ & $ 8.4\% $ &  $ 7.7\% $ & 7540 \\
1998-99 & $ 23.7 $ & $ 19.0 $ & 0.79 & $ 19.6 $ & 0.25 & $ 10\% $ & $ 11\%  $ &  $ 7.1\% $ & 7800 \\
1999-00 & $ 24.2 $ & $ 19.6 $ & 0.78 & $ 19.3 $ & 0.30 & $ 11\% $ & $ 11\%  $ &  $ 7.2\% $ & 7800 \\ \hline
\end{tabular}
\label{table}
\end{table}

However, by construction, $C(x)$ is always a monotonous function, so
one may argue that different CDFs look visually similar and hard to
distinguish.  Thus, it is instructive to consider PDF as well, which
we do in the next Section.

\section{Probability Density Function}
\label{sec:PDF}

\begin{figure}[b]
\includegraphics[width=0.49\linewidth]{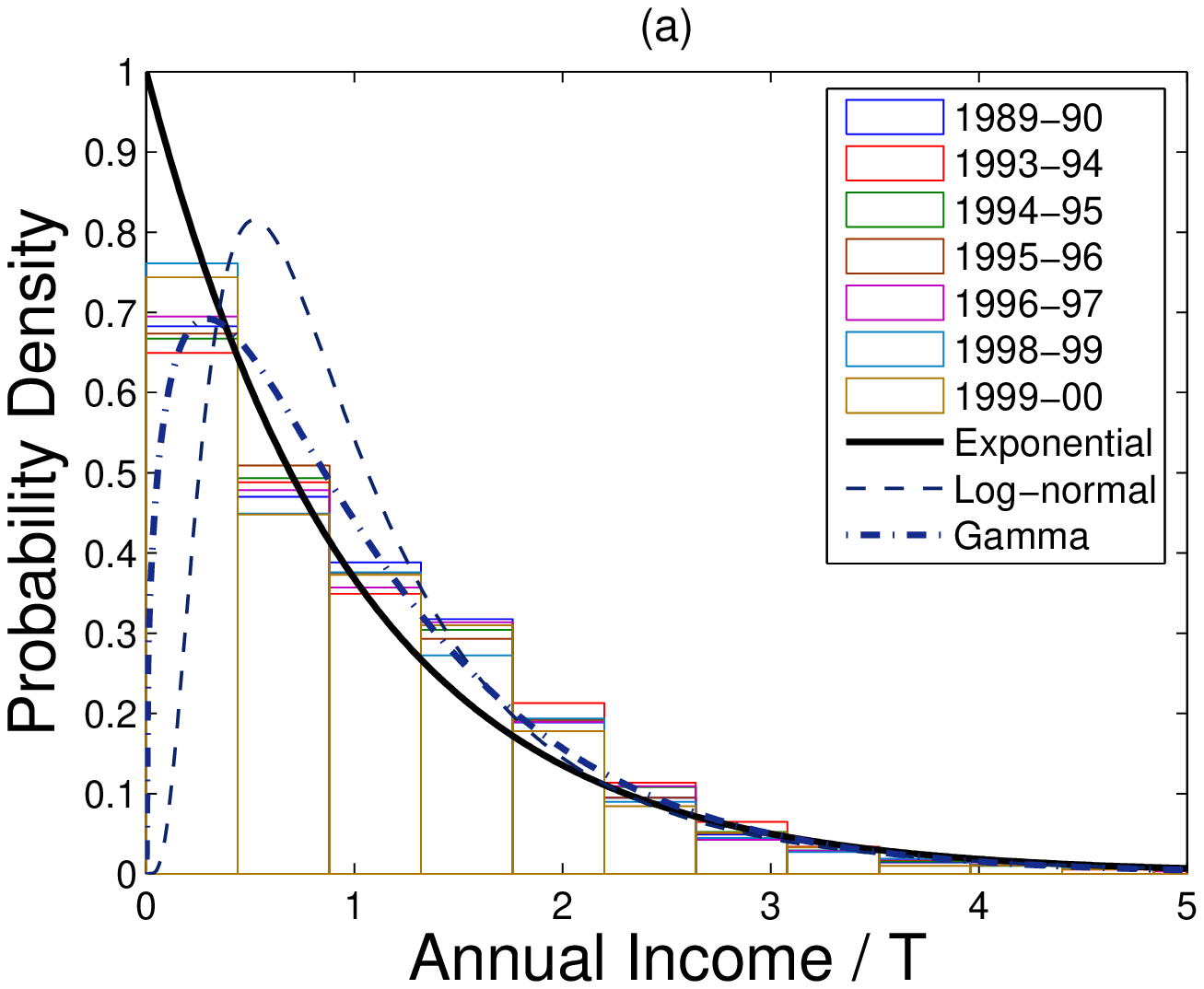}
\hfill
\includegraphics[width=0.49\linewidth]{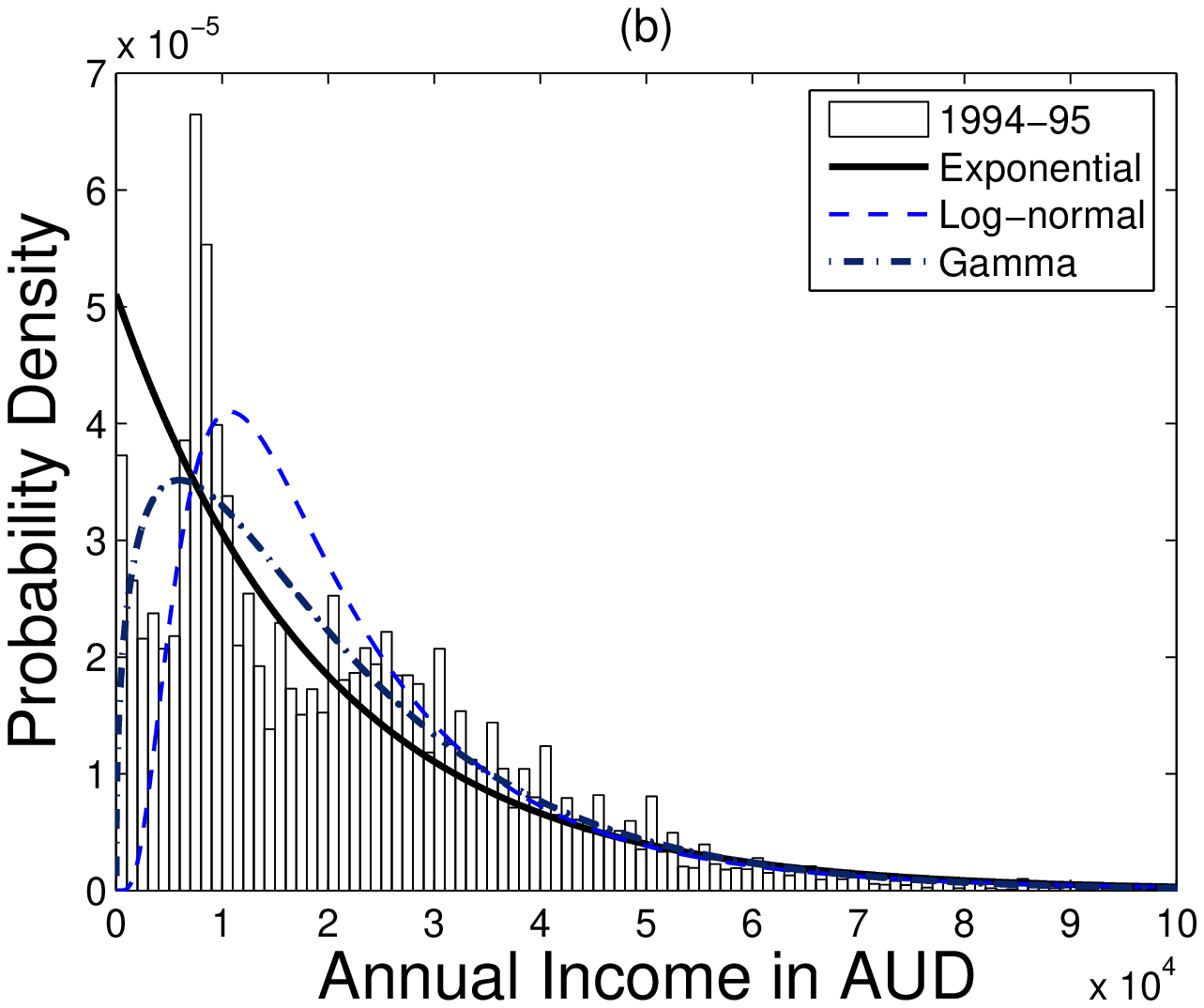}
\caption{The probability density function (PDF) of income distribution
  shown with coarse-grained (a) and high (b) resolutions.  The lines
  show fits to different theoretical functions in Eq.\ (\ref{pdf}).}
\label{fig:PDF}
\end{figure}

In order to construct $P(x)$, we divide the income axis into bins of
the width $\Delta x$, calculate the sum of the weights $w_n$ of the
individuals with incomes from $x$ to $x+\Delta x$, and plot the
obtained histogram.  However, there is subjectiveness in the choice of
the width $\Delta x$ of the bins.  If the bins are too wide, the
number of individuals in each bin is big, so the statistics is good,
but fine details of the PDF are lost.  If the bins are too narrow, the
number of individuals in each bin is small, thus relative fluctuations
are big, so the histogram of PDF becomes noisy.  Effectively, $P(x)$
is a derivative of the empirical $C(x)$.  However, numerical
differentiation increases noise and magnifies minor irregularities of
$C(x)$, which are not necessarily important when we are interested in
the universal features of income distribution.  To illustrate these
problems, we show PDFs obtained with two different bin widths in Fig.\ 
\ref{fig:PDF}.

Fig.\ \ref{fig:PDF}(a) shows the coarse-grained histogram of $P(x)$
for all years with a wide bin width $\Delta x/T\approx0.43$.  The
horizontal axis represents income $x$ rescaled with the values of $T$
from Table \ref{table}. The lines show the exponential, log-normal,
and gamma fits with the same parameters as in Fig.\ \ref{fig:CDF}.
With this choice of the bin width, the empirical $P(x)$ is a
monotonous function of $x$ with the maximum at $x=0$, and the
exponential function gives a reasonable overall fit.  The log-normal
and gamma fits have maxima at $x/T\approx0.56$ and $x/T\approx0.29$.
These values are close to the bin width, so we cannot resolve whether
$P(x)$ has a maximum at $x=0$ or at a non-zero $x$ within the first
bin.

Fig.\ \ref{fig:PDF}(b) shows the PDF for the year 1994-95 with a
narrow bin width $\Delta x=1$ k\$, which corresponds to $\Delta
x/T\approx0.05$.  This PDF cannot be fitted by any of the three
distributions, because it has a very sharp and narrow peak at the low
income 7.3~k\$, which is way below the average income of 19.6~k\$ for
this year.  This peak is present for all years, and its position is
reported in the last column of Table \ref{table}.  The peak is so
sharp and narrow that it cannot be attributed to the broad maxima of
the log-normal or gamma PDFs.  We speculate that this peak occurs at
the threshold income of some sort of government policy, such as social
welfare, minimal wage, or tax exemption.  Comparing the empirical PDF
with the exponential curve, shown by the solid line, we infer that the
probability density above and below the peak is transferred to the
peak, thus creating anomalously high population at the special income.

We also studied how often different income values occur in the data
sets.  The most frequently reported incomes for different years are
always round numbers, such as 15~k\$, 20~k\$, 25~k\$, etc.  This
effect can be seen in the periodically spaced spikes in Fig.\ 
\ref{fig:PDF}(b).  It reflects either the design of the survey
questionnaires, or the habit of people for rounding their incomes in
reporting.  In addition to the round numbers, we also find the income
corresponding to the peak position among the five most frequently
reported incomes.  This income, shown in the last column in Table
\ref{table}, is not round and changes from year to year, but sometimes
stays the same.  This again suggests that the sharp peak in Fig.\ 
\ref{fig:PDF}(b) is the result of a government-imposed policy and
cannot be explained by statistical physics arguments.

By definition, $P(x)$ is the slope of $C(x)$ with the opposite sign.
Fig.\ \ref{fig:CDF} clearly shows that the slope of $C(x)$ at $x=0$ is
not zero, so $P(x=0)\neq0$.  Fig.\ \ref{fig:PDF}(b) also shows that
the probability density at zero income is not zero.  In fact, $P(x=0)$
is higher than $P(x)$ for all other $x$, except in the narrow peak.
The non-vanishing $P(x=0)$ is a strong evidence against the
log-normal, gamma, and similar distributions, but is qualitatively
consistent with the exponential function.  However, there is also
substantial population with zero and negative income, which is not
described by any of these theories.

\section{Discussion and Conclusions}

All three functions in Eq.\ (\ref{pdf}) are the limiting cases of the
generalized beta distribution of the second kind (GB2), which is also
discussed in econometric literature on income distribution
\cite{Mcdonald}.  GB2 has four fitting parameters, and distributions
with even more fitting parameters are considered in literature
\cite{Mcdonald}.  Generally, functions with more parameters are
expected fit the data better.  However, we do not think that
increasing the number of free parameters gives a better insight into
the problem.  We think that a useful description of the data is the
one that has the minimal number of parameters, yet reasonably (but not
necessarily perfectly) agrees with the data.  From this point of view,
the exponential function has the advantage of having only one
parameter $T$ over the log-normal, gamma, and other distributions with
more parameters.  Fig.\ \ref{fig:CDF}(a) shows that $\log C$ vs.\ $x$
is approximately a straight line for about 98\% of population,
although small systematic deviations do exist.  The log-normal and
gamma distributions do not improve the fit significantly, despite
having more parameters, and actually mimic the exponential function.
Thus we conclude that the exponential function is the best choice.

The analysis of PDF shows that the probability density at zero income
is clearly not zero, which contradicts the log-normal and gamma
distributions, but is consistent with the exponential one, although
the value of $P(x=0)$ is somewhat lower than expected.  The
coarse-grained $P(x)$ is monotonous and consistent with the
exponential distribution.  The high resolution PDF shows a very sharp
and narrow peak at low incomes, which, we believe, results from
redistribution of probability density near the income threshold of a
government policy.  Technically, none of the three function in Eq.\
(\ref{pdf}) can fit the complicated, three-peak PDF shown in Fig.\
\ref{fig:PDF}.  However, statistical physics approaches are intended
to capture only the baseline of the distribution, not its fine
features.  Moreover, the deviation of the actual PDF from the
theoretical exponential curve can be taken as a measure of the impact
of government policies on income redistribution.

\bigskip {\bf Acknowledgments.}  
T.~Di Matteo wishes to thank the Australian Social Science Data
Archive, ANU, for providing the ABS data and the partial support by
ARC Discovery Projects: DP03440044 (2003) and DP0558183 (2005), COST
P10 `Physics of Risk' project and M.I.U.R.-F.I.S.R. Project
`Ultra-high frequency dynamics of financial markets'.



\end{document}